\magnification 1200
\input jnl_ref.tex
%1st updating with effect from: 5 Sept 1991

%2ND UPDATING WITH EFFECT FROM: 28 JUNE 1993
%(for the purpose of making PlainTex file + Latex file identical)

%------------------------------------------------------------------------
\headline={\ifnum\pageno=1\firstheadline\else
\ifodd\pageno\rightheadline \else\leftheadline\fi\fi}
\def\firstheadline{\hfil}
\def\rightheadline{\hfil}
\def\leftheadline{\hfil}
	\footline={\ifnum\pageno=1\firstfootline\else\otherfootline\fi}
\def\firstfootline{\rm\hss\folio\hss}
\def\otherfootline{\hfil}

\font\twelvebf=cmbx10 scaled\magstep 1
 1 
 1

\font\tenbf=cmbx10
\font\tenrm=cmr10
\font\tenit=cmti10

\font\eightbf=cmbx8
\font\eightrm=cmr8

\parindent=1.5pc
\hsize=6.0truein
\vsize=8.5truein
\nopagenumbers

%%%%%%%%%%%%%%%%%%%%%%%%%%%%%%%%%%%%%%%%%%%%%%%%%%%%%%%%%%%%%%%%%%%%%%%%%%%%%%%%%

% RAB MACROS

\def\fa{f_{\rm a}}
\def\ma{m_{\rm a}}
\def\oa{\Omega_{\rm a}}
\def\tentw{\bigg{(}{\fa\over 10^{12}{\rm GeV}}\bigg{)}^{1.18}}
\def\lskip{\vskip 5pt}

% EQUATION NUMBERING

\def\new{{\the\eqnumber}\global\advance\eqnumber by 1}
\def\delaynew{{\the\eqnumber}}
\def\nownew{\global\advance\eqnumber by 1}
\def\last{\advance\eqnumber by -1 {\the\eqnumber}
    \global\advance\eqnumber by 1}
\def\eqnam#1{%%%Naming macro
\xdef#1{\the\eqnumber}}

% MACROS

\def\lapp{\hbox{$ {     \lower.40ex\hbox{$<$}
                   \atop \raise.20ex\hbox{$\sim$}
                   }     $}  }
\def\gapp{\hbox{$ {     \lower.40ex\hbox{$>$}
                   \atop \raise.20ex\hbox{$\sim$}
                   }     $}  }

\def\marbul{\strut\vadjust{\kern-2pt$\bullet$}}

\def\rr{\rangle}
\def\ll{\langle}

\def\specialwarn{\vtop to
\strutdepth{\baselineskip\strutdepth\vss\llap{
\lower.1ex\hbox{$\bigtriangleup$}\kern-0.884em$\triangle$\kern-0.5667em{\eightrm
!}\hskip 13.5pt}\null}}
\def\strutdepth{\dp\strutbox}

\def\figure#1#2#3#4#5{
\topinsert
\null
\medskip
\vskip #2\relax
\null\hskip #3\relax
\special{illustration #1}
\medskip
{\baselineskip 10pt\noindent\narrower\rm\hbox{\eightbf
Figure #4}:\quad\eightrm
#5 \smallskip}
\endinsert}

\def\doublefigure#1#2#3#4#5#6#7{
\topinsert
\null
\medskip
\vskip #3\relax
\null\hskip #4\relax
\special{illustration #1}
\hskip #5
\special{illustration #2}
\medskip
{\baselineskip 10pt\noindent\narrower\rm\hbox{\eightbf
Figure #6}:\quad\eightrm
#7 \smallskip}
\endinsert}

\def\caption#1#2{
\baselineskip 10pt\noindent\narrower\rm\hbox{\eightbf
#1}:\quad\eightrm
#2 \smallskip}

\def\picture #1 by #2 (#3){
  \vbox to #2{
    \hrule width #1 height 0pt depth 0pt
    \vfill
    \special{picture #3} % this is the low-level interface
    }
  }

\def\scaledpicture #1 by #2 (#3 scaled #4){{
  \dimen0=#1 \dimen1=#2
  \divide\dimen0 by 1000 \multiply\dimen0 by #4
  \divide\dimen1 by 1000 \multiply\dimen1 by #4
  \picture \dimen0 by \dimen1 (#3 scaled #4)}
  }

%%%%%%%%%%%%%%%%%%%%%%%%%%%%%%%%%%%%%%%%%%%%%%%%%%%%%%%%%%%%%%%%%%%%%%%%%%%%%%%%%

\newcount\eqnumber

%%%%%%%%%%%%%%%%%%%%%%%%%%%%%%%%%%%%%%%%%%%%%%%%%%%%%%%%%%%%%%%%%%%%%%%%%%%%%%%%%

\centerline{\tenbf CURRENT STATUS OF AXION
COSMOLOGY\footnote{\dag}{To appear in the
Proceedings of the International Symposium
on the Critique of the Sources of Dark
Matter in the Universe, UCLA, 16th-18th
February 1994.\smallskip}} \vglue 0.8cm
\centerline{\tenrm R.A.Battye \&
E.P.S.Shellard} \baselineskip=13pt
\centerline{\tenit D.A.M.T.P., University of
Cambridge } \baselineskip=12pt
\centerline{\tenit Silver Street, Cambridge,
CB3 9EW, England} \vglue 0.8cm
\centerline{\tenrm ABSTRACT} \vglue 0.3cm
{\rightskip=3pc\leftskip=3pc\eightrm\baselineskip=10pt\noindent
We review the current status of axion
cosmology in the light of recent work. We
conclude that in the standard scenario in
which a global string network forms at the
Peccei-Quinn phase transition, the largest
contribution to the axion density is from
string loops, even when uncertainties in the
nature of the QCD phase transition are taken
into account. We also briefly discuss the
possible  implications of non standard
scenarios, such as an inflationary epoch and
entropy production.  \vglue 0.6cm}

\rm
\baselineskip=11pt
\eqnumber=1

\leftline{\twelvebf 1. Introduction}
\vglue 0.4cm
\noindent The axion was first proposed as an elegant solution to the strong CP
problem of the standard model of particle physics\refto{PQ}. However, it was
soon realised that it could have an important role to play as a dark matter
candidate, since it acquires a small mass at the QCD phase
transition\refto{WW}. Thus, the cosmology of the axion has attracted considerable
interest, since requiring that the axion density is below critical ($\oa<1$)
provides an upper bound on the Peccei-Quinn symmetry breaking scale
$\fa$. Combined with constraints from  accelerator searches and
astrophysics\refto{SN} ($\fa\gapp 10^{9-10}$GeV), one can tightly constrain the
properties of the axion.

The earliest estimates of the axion density\refto{Hom} assumed that axions were
produced through coherent oscillations about the minimum of the potential due to
homogeneous misalignments of the axion field for $T>\Lambda_{\rm QCD}$. The most
accurate estimate of the relative contribution of these zero momentum axions is
\eqnam{\homcont} $$\Omega_{{\rm a},{\rm h}}  \approx 0.9 h^{-2}\Delta
\bigg{(}{\fa\over 10^{12}{\rm GeV}}\bigg{)}^{1.18}  \bar\theta^2_i \,,\eqno(\new)$$
where $\Delta$ accounts for the uncertainties in the QCD phase
transition\refto{Turn,BSb}, \eqnam{\uncertainty} $$\Delta= 10^{\pm
.5}\bigg{(}{\bar\ma\over 6\times 10^{-6}{\rm
eV}}\bigg{)}^{0.82}\bigg{(}{\Lambda_{\rm QCD}\over 200{\rm
MeV}}\bigg{)}^{0.65}\bigg{(}{{\cal N}_{\rm QCD}\over 60}\bigg{)}^{-0.41}
\,,\eqno(\new)$$  and Hubble's constant at the present day is $H_{0}= 100h\,{\rm
km}\, {\rm s}^{-1}\,{\rm Mpc}^{-1}$, $0.35<h<1.0$. Assuming the average value of
the axion field $\langle \bar\theta^2_i\rangle = \pi/3$, implies a constraint
$\fa\lapp 10^{12}{\rm GeV}\,,\ma\gapp 5\mu {\rm eV}\,.$

However, this estimate ignored topological effects at the Peccei-Quinn
phase transition, that is the production of a network of global strings formed
by the Kibble mechanism\refto{VE}.
The implications of the decay of this global string network has been the subject of
much controversy\footnote{\dag}{We note that this controversy may be avoided
completely if the reheat temperature of any inflationary epoch is lower than
$\fa$ (see \S4).}, in particular regarding  the nature of the radiation spectrum of
the strings\refto{Dav,DSa,Sika,Sika1}. In a recent publication\refto{BSa}, we
investigated this issue, constructing a tight mathematical argument passing from
the Goldstone model with global strings through to the Kalb-Ramond action with
antisymmetric tensors. We found that analytic radiation calculations using the
antisymmetric tensor formalism were in good quantitative agreement with the
numerical simulations of global strings in the Goldstone model. These findings
were in broad agreement with the original work of Davis {\it et al}$\,$
\refto{Dav,DSa} and contrary to the predictions of Sikivie {\it et al}
$\,$\refto{Sika,Sika1}. Taken at face value these results support a constraint which
would rule out the axion in the standard scenario. However, we also noted that,
although the underlying physics of the original work was correct, the model
employed for string evolution was too simplistic\refto{She,VS}. This work was
expanded on in a recent letter\refto{BSb} which, subject to several uncertainties,
favoured a constraint of $\fa\lapp 10^{11}{\rm GeV}\,,\ma\gapp 50\mu {\rm
eV}\,.$

\vglue 0.6cm
\leftline{\twelvebf 2. The nature of global string dynamics}
\vglue 0.4cm

\noindent The Kalb-Ramond action for global strings can be easily deduced in the
low energy limit from the Goldstone action, using the massive duality 
relation $\phi^2\partial_{\mu}\vartheta=\textstyle{1\over
2}\fa\epsilon_{\mu\nu\lambda\rho}\partial^{\nu}B^{\lambda\rho}\,$\refto{DSb}, 
\eqnam{\kraction} $$ S=-\mu_0\int d\sigma d\tau\sqrt{-\gamma} - {1\over 6}\int
d^4x H^2 - 2\pi\fa\int B_{\mu\nu}d\sigma^{\mu\nu}\,,\eqno(\new)$$ and its
relevance even at higher energies has been established numerically in
ref.[12]. The similarity of this action to the Nambu action used to describe
gauge strings, would tend to imply that the dynamics of global strings is
qualitatively the same, that is:

\lskip
\noindent (i) A Brownian network of strings will evolve towards a
self similar scaling solution in which all the properties of the string network
remain constant with respect to the horizon\refto{ASa}. In particular the
density of long strings per horizon, $\rho_{\infty}= \zeta\mu/t^2$, and the loop
production size with respect to the horizon, $\ll l\rr=\alpha t$, will remain
constant\refto{BSb,She}.

\lskip
\noindent (ii) An initially perturbed string oscillating relativistically will
radiate with a classical spectrum dominated by the second fundamental mode, that is the
radiation pattern will be quadropole. Fig.1 illustrates the pattern of
radiation for a global string from numerical field theory simulations and the
fall off of the amplitude\refto{BSa}. Note that the amplitude is not critically
damped as suggested in refs.\refto{Sika,Sika1}.   

\doublefigure{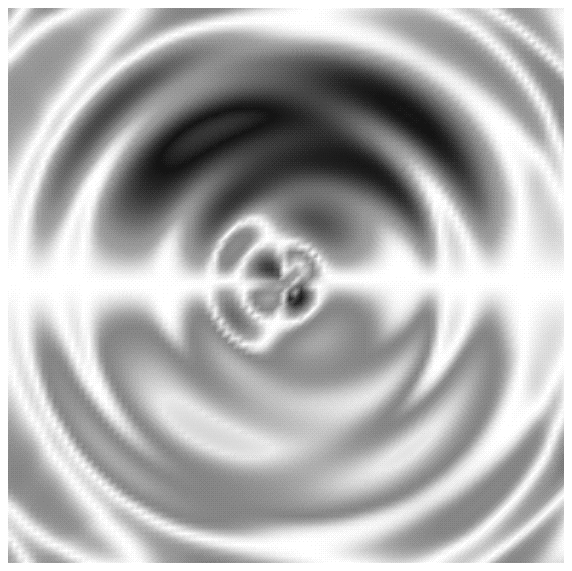}{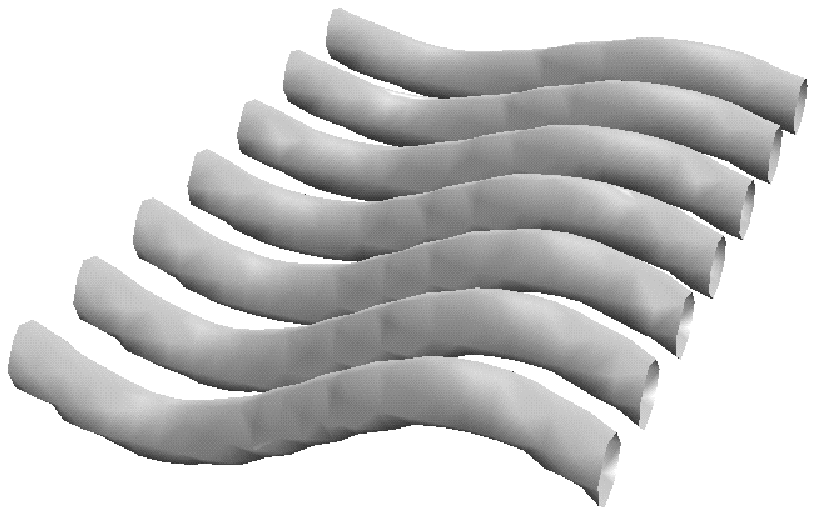}{1.8in}{-0.2in}{2.1in}{1} {(a)
Radiation quadropole pattern in the plane perpendicular to the string (b)
Isosurfaces of the energy density for an evolving sinusoidal perturbation  shown
after each period.  Radiation damping causes a noticeable decrease in the
oscillation amplitude.}

\lskip
\noindent (iii) The radiation power from long strings is generically proportional
to $\varepsilon^4/\lambda$, where $\varepsilon = 2\pi A/\lambda$, $A$ is the
amplitude of the perturbation and $\lambda$ is the wavelength\refto{BSb}, whereas
the radiation power from loops is independent of the size of the
loop\refto{VVa}. 

\lskip
\noindent However, there are some important differences due to the topological
coupling of the Goldstone boson field to the string and non local nature of the
self field: 

\lskip
\noindent (i) The global strings decay primarily into Goldstone bosons (or
axions), as opposed to gravitational radiation, via an enhanced radiation
mechanism. This radiation is a number of orders of magnitude stronger than
gravitational radiation, leading to  different properties for the string network on
small scales. In particular, small scale structure will be removed much more
efficiently. As a consequence of this the loop production size with respect to
the horizon can be expected to be considerably larger.

\lskip
\noindent (ii) Global strings appear to be non local objects, with a
logarithmically  divergent energy per unit length. However, within both the
Goldstone and Kalb-Ramond formalisms this can be renormalised using the inter
string separation $R$, and the core radius $\delta$, as renormalisation cut offs. In
the Kalb-Ramond formalism the self field is\refto{DQ,BSc} 
$ f^{\mu}_{\rm self} = 2\pi\fa^2\log(R/\delta)(\ddot X^{\mu} -
X^{\prime\prime\mu})+O(R^2)$ and the renormalised string tension is  
$\mu({R})=\mu_0+2\pi\fa^2\log(R/\delta)\,.$   This implies that the
self field of the string moves with the string core and that the radiation field
can be calculated, to a reasonable degree of accuracy, by subtracting the static
self field when the string is straight in the Goldstone formalism. 

\vglue 0.6cm
\leftline{\twelvebf 3. Contributions to the axion density from 
defects}
\vglue 0.4cm

\noindent To describe scale-invariant loop creation and decay we must define
several further parameters: Firstly, the loop
backreaction parameter $\kappa$ describes the radiation power from loops which is
given by\refto{VVa}
\eqnam{\looppower}
$$P=\Gamma_{\rm a}\fa^2=\kappa\mu\,,\eqno(\new)$$
where $\Gamma_{\rm a}$ is a factor dependent on the loop trajectory, but not 
its size, which is estimated to be $\langle \Gamma_{\rm a} \rangle 
\approx 65$\refto{VVa,QSSP,ASb} (exploiting similarities with gravitational
radiation).   As the loop decays into axions, its
radius shrinks linearly 
$\ell=\ell_{\rm i}-\kappa(t-t_{\rm i})\,,$
where $t_{\rm i}$ is the loop creation time and $\ell_{\rm i}=\ell(t_{\rm i})$.
Typically, in a cosmological context we have 
$\kappa \approx (\Gamma_{\rm
a}/2\pi)[\ln(t/\delta)]^{-1}\approx 0.15$.
Secondly, we define the long string backreaction parameter $\gamma$; length scales
below $\gamma t$ are smoothed by radiative damping in one Hubble time.  
The study in ref.[12]
indicated that long string radiation
was somewhat weaker than that for loops with $\gamma\sim 0.1\kappa$.  The
significance of $\gamma$ is that it should set the minimum loop creation size,
that is, we expect $\gamma\lapp \alpha\lapp\kappa$. 
Given these assumptions and energy conservation
considerations, our scale-invariant model implies that the number density of loops 
in the interval $\ell$ to $\ell +d\ell$ is given by
\eqnam{\numdenloop} 
$$
n(\ell,t)d\ell={\nu d\ell\over
t^{3/2}(\ell+\kappa t)^{5/2}}\,,\qquad \ell \le \alpha t\,.\eqno(\new)
$$

The energy density of massless axions emitted at time $t_1$ in an interval $dt_1$
with frequencies from $\omega_1$ to $\omega_1+d\omega_1$ is $$ d\rho_{\rm
a}(t_1)=dt_1d\omega_1\fa^2\int_{0}^{\infty}d\ell\,n(\ell,t_1)\ell\,g(\ell\omega)\,,
\eqno(\new) $$ where $g(x)$ is a function normalized upon integration by
$\Gamma_a$. Assuming ${\cal N}$ constant, the spectral density can be calculated by
integrating  over $t_1<t$, taking into account the redshifting of both the
frequency, $\omega = {a(t_1)/ a(t)}\omega_1$, and the energy density, $\rho_{\rm
a}\propto a^{-4}$.  Neglecting the slow logarithmic dependence of the backreaction
scale $\kappa$ and truncating $g(x)$ we have
 \eqnam{\approxspecdensity} $${d\rho_{\rm a}\over d\omega}(t)\approx {4\Gamma_{\rm
a}\fa^2\nu \over 3\omega
\kappa^{3/2}t^2}\left[1-\left(1+{\alpha\over\kappa}\right)^{-3/2}\right]\,.\eqno(\new)$$

From this expression we can obtain the spectral number density and upon integration
the axion number density to entropy ratio, with $s=2\pi^2{\cal N}T^3/45$.
The key epoch to estimate this ratio is at the time of wall domination $t_{\rm
w}$, when axion production by the network ends abruptly (see below), yielding 
\eqnam{\loopnument} $${n_{\rm a}\over s}\approx 6.7\times10^6 \left[1
 - \left(1+{\alpha\over \kappa}\right)^{-3/2}\right]\Delta\bigg{(}{\bar\ma\over
6\times 10^{-6}{\rm eV}}\bigg{)}^{-1}\bigg{(}{\fa\over 10^{12}{\rm
GeV}}\bigg{)}^{2.18}\,,$$ where we have taken typical parameter values $\Gamma_{\rm
a}\approx 65$, $\nu\approx 0.40\zeta\alpha^{1/2}$ and $\zeta \approx 13$. Assuming
number conservation after this time and using the entropy density
$s_{0}=2809(T_{0}/2.7{\rm K})^3{\rm cm}^{-3}$ and critical density $\rho_{\rm
crit}=1.88\times10^{-29}h^2{\rm g cm}^{-3}$ at the present day, one can deduce 
that the axion loop contribution is  \eqnam{\loopcont} $$ \Omega_{\rm
a,\ell}\approx 10.7 \left(\alpha\over\kappa\right)^{3/2}
\left[1-\left(1+{\alpha\over\kappa}\right)^{-3/2}\right]
h^{-2}\Delta\bigg{(}{T_{0}\over 2.7{\rm K}}\bigg{)}^{3}\bigg{(}{\fa\over
10^{12}{\rm GeV}}\bigg{)}^{1.18}\,.\eqno(\new) $$ 

Assuming radiative dominance of the smallest scale $\gamma t$ for the long string
network (as observed in ref.[20]), one can calculate its contribution to the axion
density, \eqnam{\longcont} $$\Omega_{{\rm a},\infty}\approx
1.2h^{-2}\Delta\bigg{(}{T_0\over 2.7{\rm K}}\bigg{)}^3\tentw\,.\eqno(\new)$$
However, the considerable uncertainty of Eq.(\longcont) must be emphasized given
its sensitivity to the amplitude of small scale structure and the overall string
radiation spectrum. Fig. 2 summarizes the best estimates of constraint on the
symmetry breaking scale $\fa$ for various values of $\alpha/\kappa$, assuming the
only contribution to the axion density comes from strings.

\figure{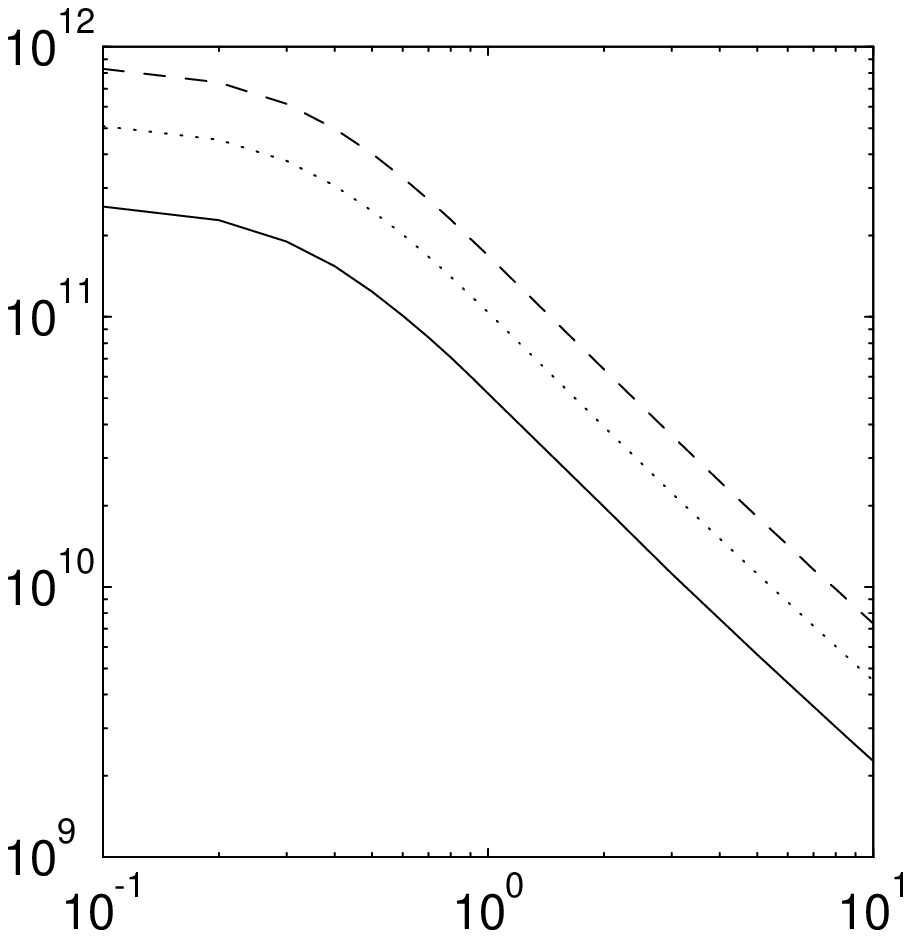}{3.5in}{0.4in}{2}{The constraint on the symmetry breaking scale
$\fa$ for various values of $\alpha/\kappa$, assuming the only contribution to the
axion density comes from stings, that is the sum of Eqs. (\loopcont) and
(\longcont). The solid line corresponds to $h=0.5$, the dotted line to $h=0.75$ and
the dashed line to $h=1.0$}

Near the QCD phase transition the axion
acquires a mass and the network evolution alters dramatically because domain walls
form\refto{Sikb}, with each string becoming attached to a wall\refto{VE}. Large
field variations due to the strings collapse into domain walls and subsequently the
domain walls dominate the string dynamics. The demise of the string--wall network
proceeds rapidly\refto{VE}, as demonstrated numerically\refto{She,RPS}. The strings
frequently intersect and intercommute with the walls, effectively `slicing up' the
network into small oscillating walls bounded by string loops. Multiple
self-intersections will reduce these pieces in size until the strings dominate the
dynamics again and decay continues through axion emission. The contribution to the
axion density from this decay has been estimated as\refto{Lyth1,Jap} 
\eqnam{\domaincont} $$\Omega_{{\rm a},{\rm w}} \sim
O(1)h^{-2}\Delta\bigg{(}{T_0\over 2.7{\rm K}}\bigg{)}^3\bigg{(}{\fa\over
10^{12}{\rm GeV}}\bigg{)}^{1.18}\,.\eqno(\new)$$

\vglue 0.6cm
\leftline{\twelvebf 4. Non-standard scenarios}
\vglue 0.4cm

\noindent The discussion of the axion density presented in \S3 presumes that a
global string network forms after any epoch of inflation, that is $T_{\rm
reh}>\fa$. In the alternative scenario when $T_{\rm reh}<\fa$, the axion density
and any strings formed before inflation are exponentially
suppressed\footnote{\dag}{ It is possible to form strings during
inflation\refto{Lyth2}, however the significance of this possibility is far from
clear at the present, since no quantitative predictions have been made to date.}. In
this case, the only contribution to the axion density comes from the initial
misalignment mechanism, that is $\oa$ is given by Eq. (\homcont). However,
$\theta_{\rm i}$ is homogeneous on scales larger then the current horizon  and
there is no a priori reason to suppose that $\theta_{\rm i}$ should take any
particular value. Essentially, with the freedom to choose any value of $\theta_i$,
there is no constraint on $\fa$. Some authors\refto{Lin} seem to favour larger
values of $\theta_{\rm i}(\sim \pi/2)$ as being more natural, since they avoid
apparent fine tuning and anthropic arguments about our region of the universe. If
one conservatively constrains $\theta_{\rm i}$ to be in the range
$0.1\lapp\theta_{\rm i}\lapp \pi /2$ the constraint on the axion is $\fa\lapp
10^{11}-10^{14}\,,\ma\gapp 0.05-50\mu{\rm eV}$.

Various attempts have been made to avoid the anthropic arguments described above by
appealing to particle physics motivated models of inflation in which $\theta_{\rm
i}$ is set by the conditions at the end of inflation, possibly the best so far being
hybrid inflation\refto{LinHy}. In these models the axion field is coupled to the
inflaton causing the formation of topological defects at the end of inflation. It
has been suggested that these models occur naturally in supersymmetric axion
models\refto{false}.

Entropy production by the out of equilibrium decay  of massive particles between
the QCD phase transition and nucleosynthesis, can weaken all the bounds on $\fa$.
If the entropy is increased by some factor $\beta$, that is $s\rightarrow \beta s$,
then the axion density is decreased by a factor $\beta^{-1}$, that is
$\Omega_{\rm a}\rightarrow \beta^{-1}\Omega_{\rm a}$  For example, it has been
noted that that the decay of the saxino --  the spin zero partner of the axion --
can lead to a dilution by\refto{Lyth3,Kim}  \eqnam{\decay} $$\beta < 5\times
10^{3}\bigg{(}{m_{\rm sax}\over 1{\rm TeV}}\bigg{)}\,,\eqno(\new)$$ which can be
up to 1000.

\vglue 0.6cm
\leftline{\twelvebf 5. Discussion}
\vglue 0.4cm

\noindent At present the best motivated scenario for the axion is one in which a
network of global strings forms at the Peccei-Quinn phase transition.
The understanding of the evolution of this network has been clouded by some 
misunderstandings as to the nature of global string dynamics. Here we have
summarized the results of our recent work which we believe confirms the growing
consensus which has emerged in the literature \refto{Dav,DSa,DQ,VVa,VS}. 
 
However, we have also noted that the constraint from strings is somewhat
weaker than previously thought. Comparing
Eqs.(\loopcont),(\longcont) and (\domaincont) implies that the contribution from 
string loops is the largest for $\alpha/\kappa\gapp 0.45$, irrespective of the
values of $h,\Delta$ and $T_0$. The ratio $\alpha/\kappa$ expresses the key
uncertainty arising from our inadequate understanding of long string radiative
backreaction. In the case of $\alpha/\kappa\approx 1$,  the axion density is
dominated by the loop contribution and the constraint is 
\eqnam{\conone}
$$\fa \lapp 6.0\times 10^{10}{\rm GeV}\,,\quad \ma\gapp 100\mu{\rm eV}\,,\quad
h=0.5\eqno(\new)$$
or $\fa\lapp 1.9\times 10^{11}{\rm GeV}\,,\ma\gapp 31\mu{\rm eV}$, for $h=1.0$
(we have not included the parameter uncertainties of Eq.(\uncertainty)), whereas if
$\alpha/\kappa \approx 0.1$ the constraint is less certain since Eq.(\longcont) and
Eq.(\domaincont) are more important. In this case we quote the far less accurate
constraint,
\eqnam{\contwo}
$$\fa\lapp 5.0 \times 10^{11}{\rm GeV}\,,\quad\ma\gapp 10\mu{eV}\,.\eqno(\new)$$
In each case there is still a narrow window for the axion to exist, although
marginally outside the range of current experimental searches\refto{vanbib}.

Nonetheless, we have noted that there are other non-standard scenarios in which a
lower mass axion can exist, such as models with a low reheat inflationary epoch or
with entropy production. We have briefly summarized these scenarios in \S4, although
the uncertainties in these scenarios must be stressed. The cosmology of the axion
remains  fruitful area for further research.

\vglue 0.6cm
\leftline{\twelvebf 6. Acknowledgements}
\vglue 0.4cm

\noindent We are grateful for helpful discussions with Alex Vilenkin, David Lyth,
Georg Raffelt, Scott Thomas, Atish Dabholkar, Brandon Carter
and Michael Turner. We both acknowledge the support of the Science and Engineering
Research Council, in particular the Cambridge Relativity group rolling grant
(GR/H71550) and Computational Science Initiative grants (GR/H67652 \& GR/H57585)

\vglue 0.6cm
\leftline{\twelvebf 7. References}
\vglue 0.4cm

%%%%%%%%%%%%%%%%%%%%%%%%%%%%%%%%%%%%%%%%%%%%%%%%%%%%%%%%%%%%%%%%%%%%%%%%%%%%%%%%%%
% References
%%%%%%%%%%%%%%%%%%%%%%%%%%%%%%%%%%%%%%%%%%%%%%%%%%%%%%%%%%%%%%%%%%%%%%%%%%%%%%%%%%

%%% PARAGRAPH SHAPE:

%\def\hang{\hangindent20pt\hangafter1\noindent}
\def\hang{}

%%% STANDARD JOURNAL:
%%% The following information must be supplied.
%%% \jnl{Authors|year|Article title|Journal|Volume|Page} 

\def\jnl#1#2#3#4#5#6{\hang{#1, {\it #4\/} {\bf #5} (#2) #6.}
									}

%%% JOURNAL WITH ERRATUM:
%%% \jnlerr{Authors|year|Article title|Journal|Volume|Page|Err Volume|Err page} 

%%% TWO JOURNALS:
%%% \jnltwo{Authors|year|Article title|Journal|Volume|Page|Journal|Vol.|Page} 

\def\jnltwo#1#2#3#4#5#6#7#8#9{\hang{#1  {\it #4\/} {\bf #5}(#2) #6;
{\it #7\/} {\bf #8}(#2) #9.}
									}

%%% PREPRINT:
%%% \prep{Authors|year|Article title|Number and comments} 
\def\prep#1#2#3#4{\hang{#1 (#2) #4.}
									}

%%% PROC:
%%% \proc{Authors|year|Article title|Book title|Editor|Publisher, City} 

\def\procu#1#2#3#4#5#6{\hang{#1 [#2], in {\it #4\/}, #5, ed.\ (#6).}
}

%%% BOOK:
%%% \book{Authors|year|Book title|Publisher, City (where applicable)} 
\def\book#1#2#3#4{\hang{#1  {\it #3\/} (#4), #2.}
									}

%%% GENERAL REF:
%%% \genref{Author|year|Whatever} 

%%% Commonly used journals

\def\prl{Phys.\ Rev.\ Lett.}
\def\pr{Phys.\ Rev.}
\def\pl{Phys.\ Lett.}
\def\np{Nucl.\ Phys.}
\def\prp{Phys.\ Rep.}

\def\apj{Ap.\ J.}

\def\cup{Cambridge University Press}

\def\skip{\vskip -4pt}

\references

\baselineskip 9pt
\let\it=\nineit
\let\rm=\ninerm
\let\bf=\ninebf
\rm 

\refis{PQ}
\jnltwo{Peccei, R.D., \& Quinn, H.R.}{1977}{CP conservation and the 
presence of pseudoparticles}{\prl}{38}{1440}{\pr}{D16}{1791}
\skip

\refis{WW}
\jnl{Weinberg, S.}{1978}{A new light boson?}{\prl}{40}{223}
\jnl{Wilczek, F.}{1978}{Problem of strong $P$ and $T$ invariance 
in the presence of instantons}{\prl}{40}{279}
\skip

\refis{VE}
\jnl{Vilenkin, A., \& Everett, A.E.}{1982}{Cosmic strings and domain walls in
models with Goldstone and pseudo-Goldstone bosons}{\prl}{48}{1867}
\skip

\refis{SN}
\jnl{Raffelt G.}{1990}{}{\prp}{198}{1} \jnl{Turner M.}{1990}{}{\prp}{197}{678}
\skip

\refis{Turn}
\jnl{Turner M.}{1986}{Cosmic and Local Mass Density of Axions}{\pr}
{D33}{889}
\skip

\refis{Lin}
\jnl{Linde, A.D.}{1991}{Axions in inflationary cosmology}{\pl}{259B}{38}
\jnl{Linde, A.D.}{1998}{Inflation and axion cosmology}{\pl}{201B}{437}
\skip

\refis{Jap}
\prep{Kawasaki, M., \& Nagaswa, M.}{1994}{Collapse of axionic domain wall and
axion emission}{University of Tokyo preprint}
\skip

\refis{LinHy}
\jnl{Linde, A.D.}{1994}{Hybrid Inflation}{\pr}{D49}{748}
\skip

\refis{Lyth1}
\jnl{Lyth, D.H.}{1992}{Estimates of the cosmological axion
density}{\pl}{B275}{279} \skip

\refis{Lyth2}
\jnl{Lyth, D.H., \& Stewart, E.D.}{1992}{Axions and Inflation : String
formation}{\pr}{D46}{532} \skip

\refis{Lyth3}
\jnl{Lyth, D.H.}{1993}{Dilution of cosmological densities by saxino
decay}{\pr}{D48}{4523}
\skip

\refis{QSSP}
\jnl{Press, W.H. , Quashnock, J.M, Scherrer, R.J., \& Spergel,
D.N.}{1991}{}{\pr}{D42}{1908}

\refis{false}
\prep{Copeland, E.J., Liddle, A.R., Lyth, D.H., Stewart, E.D. \& Wands, D.}{1994}
{False vacuum inflation}{University of Lancaster preprint}
\skip

\refis{Kim}
\jnl{Kim, J.E}{1991}{}{\prl}{67}{3465}
\skip

\refis{vanbib}
\prep{Van Bibber, K.}{1994}{}{These proceedings}
\skip

\refis{DQ}
\jnl{Dabholkar, A., \& Quashnock, J.M.}{1990}{}{\np}{B333}{815}
\skip

\refis{BSa}
\prep{Battye, R.A., \& Shellard, E.P.S.}{1994}{Global string radiation}{to appear
in {\it \np} {\bf B}}
\skip

\refis{BSb}
\prep{Battye, R.A., \& Shellard, E.P.S.}{1994}{Axion String
Constraints}{Submitted to Phys. Rev. Lett} \skip

\refis{BSc}
\prep{Battye, R.A., \& Shellard, E.P.S.}{1994}{}{In preparation} \skip

\refis{ASa}
\jnl{Bennett, D.P., \& Bouchet, F.R.}{1990}{High resolution simulations of cosmic
string evolution: network evolution}{\pr}{D41}{2408}\jnl{Allen, B., \& Shellard, E.P.S.}{1990}{Cosmic string evolution---a
numerical simulation}{\prl}{64}{119}
\skip

\refis{ASb}
\jnl{Allen, B., \& Shellard, E.P.S.}{1992}{Gravitational radiation from a
cosmic string network}{\pr}{D45}{1898}
\skip

\refis{She}
\procu{Shellard, E.P.S.}{1986}{Axionic domain walls \& cosmology}{{\rm Proceedings
of the 26th Liege International Astrophysical Colloquium,} The Origin and Early
History of the Universe}{Demaret, J.}{University de Liege}
\skip

\refis{RPS}
\jnl{Ryden, B.S., Press, W.H., \& Spergel, D.N.}{1990}{The evolution 
of networks of domain walls and cosmic strings}{\apj}{357}{293}
\skip

\refis{Sikb}
\jnl{Sikivie, P.}{1982}{Axions, domain walls and the early
universe}{\prl}{48}{1156}
\skip

\refis{Hom}
\jnl{Abbott, L.F., \&  Sikivie, P.}{1983}{A cosmological bound on the
invisible axion}{\pl}{120B}{133}
\jnl{Preskill, J., Wise, M.B., \& Wilczek, F.}{1983}{Cosmology of the invisible
axion}{\pl}{120B}{127}
\skip

\refis{Sika}
\jnl{Harari, D., \& Sikivie, P.}{1987}{On the evolution of global strings 
in the early universe}{\pl}{195B}{361}
\skip

\refis{Sika1}
\jnl{Hagmann, C., \& Sikivie, P.}{1991}{Computer simulations of the motion and
decay of global strings}{\np}{B363}{247}
\skip

\refis{Dav}
\jnl{Davis, R.L.}{1985}{Goldstone bosons in string models of galaxy
formation}{\pr}{D32}{3172}
\jnl{Davis, R.L.}{1986}{Cosmic axions from cosmic strings}{\pl}{180B}{225}
\skip

\refis{DSa}
\jnl{Davis, R.L., \& Shellard, E.P.S.}{1989}{Do axions need
inflation?}{\np}{B324}{167}
\skip

\refis{DSb}
\jnl{Davis, R.L., \& Shellard, E.P.S}{1988}{}{\pl}{214B}{1988}{219}
\skip

\refis{VVa}
\jnl{Vilenkin, A., \& Vachaspati, T.}{1987}{Radiation of Goldstone bosons from
cosmic strings}{\pr}{D35}{1138}
\skip

\refis{VS}
\book{Vilenkin, A. \& Shellard, E.P.S.}{1994}{Cosmic strings and other topological
defects}{\cup~ ({\it in press})}
\skip

\endreferences

\bye